# THEORY FOR THE DIRECT DETECTION OF SOLAR AXIONS BY COHERENT PRIMAKOFF CONVERSION IN GERMANIUM DETECTORS

R. J. Creswick, F. T. Avignone III, and H. A. Farach
University of South Carolina
Columbia, South Carolina 29208 USA

J. I. Collar
CERN, CH-1211
Geveva, 23 Switzerland

A. O. Gattone
Department of Physics
TANDAR Laboratory, C. N. E. A.
Buenos Aires, Argentina

S. Nussinov
Tel Aviv University
Ramat Aviv, Tel Aviv, Israel

K. Zioutas
University of Thessaloniki, GR-54006
Thessaloniki, Greece

Abstract

It is assumed that axions exist and are created in the sun by Primakoff conversion of photons in the Coulomb fields of nuclei. Detection rates are calculated in germanium detectors due to the coherent conversion of axions to photons in the lattice when the incident angle fulfills the Bragg condition for a given crystalline plane. The rates are correlated with the relative positions of the sun and detector yielding a characteristic recognizable sub-diurnal temporal pattern. A major experiment is proposed based on a large detector array.

Introduction

The putative pseudo-goldstone axion [1,2] had its inception almost twenty years ago following the suggested solution to the strong CP problem by Peccei and Quinn [3]. This prompted many theoretical investigations and experimental searches.

An extensive review of axion phenomenology, and their effects on stellar processes, and bounds, was given by Raffelt [4]. A detailed treatment of solar axions, and of a proposed method of detecting them, was given by van Bibber et al. [5]. In this letter we analyze a completely different direct-detection technique based on coherent Primakoff conversion of axions to photons in the lattice of a germanium detector when the incident angle satisfies the Bragg condition. Some of the basic concepts have been presented earlier in different contexts [6,7].

The axion-photon interaction Lagrangian is $L_{int} = (1/4\,M)\, a\, F_{\mu\nu} F^{\alpha\beta} \epsilon_{\mu\nu\alpha\beta}$, where "a" is the pseudoscalar axion field, $F_{\mu\nu}$ is the electromagnetic field, and $1/M \equiv g_{a\gamma\gamma}$ is the axion-photon coupling. In the Born approximation the differential cross section for elastic axion-to-photon conversion off an atom of charge Z (see Fig. 1) is:

$$\frac{d\sigma}{d\Omega} = \frac{g_{a\gamma\gamma}^2}{16\pi^2} F_a^2(2\theta) \sin^2 2\theta, \qquad (1)$$

where $2\theta$ is the scattering angle and $F(2\theta)$ is the atomic form factor. We have independently verified this cross section given previously [6,7]. Note that the axion-to-photon cross section is twice as large as that for photon-to-axion conversion because the photon has two helicities.

The form factor, $F_a(2\theta)$, is that used in earlier work [7] and can be written in two forms:

$$F_a(2\theta) = \frac{Zek^2}{r_0^{-2} + 2k^2(1-\cos 2\theta)} = \frac{Zek^2}{r_0^{-2} + q^2} = F_a(q,k), \qquad (2)$$

where $r_0$ is the screening length, and $q = 2k \sin\theta$ is the momentum exchange. The differential cross section can then be rewritten as follows:

$$\frac{d\sigma}{d\Omega} = \left(\frac{Z^2 \alpha \hbar^2}{16\pi M^2 c^2}\right) \frac{q^2(4k^2 - q^2)}{\left(r_0^{-2} + q^2\right)^2}. \qquad (3)$$

Averaging equation (3) over all solid angle yields:

$$\bar{\sigma}(\eta) = \frac{1}{4\pi} \int \frac{d\sigma}{d\pi} d\Omega = \frac{Z^2 \alpha \hbar^2}{8\pi M^2 c^2} \left[\frac{2\eta^2 + 1}{4\eta^2} \ln(1 + 4\eta^2) - 1\right], \qquad (4)$$

where $\eta \equiv r_0 k$, and $Z^2 \alpha \hbar^2 / 8\pi M^2 c^2 = 1.15 \times 10^{-44}$ cm$^2 \equiv \sqrt{\lambda}\,\sigma_0$ for $Z = 32$, where $\sqrt{\lambda} = (g_{a\gamma\gamma} \times 10^8 \text{ GeV})^2$ is the (unknown) relative strength of the axion-photon coupling.

The cross section for $\lambda = 1$ integrated over the energy range 0 to 12 keV is $1.46 \times 10^{-43}$ cm$^2$.

The expected flux of axions from the sun has been calculated by van Bibber et al. [5]. The spectrum is continuous with a peak near 4 keV and falls off exponentially beyond about 10 keV. The results of these detailed numerical calculations are well approximated by the empirical form:

$$\frac{d\Phi}{dE} = \sqrt{\lambda}\,\frac{\Phi_0}{E_0}\,\frac{(E/E_0)^3}{e^{E/E_0}-1}, \qquad (5)$$

where $\Phi_0 = 5.95 \times 10^{14}$ cm$^{-2}$ sec$^{-1}$. We have used this same expression but have adjusted the parameter $E_0 = 1.103$ keV to take into account a small change in the solar model core temperature when helium and metal diffusion are included [8].

For very light solar axions the Primakoff process in a periodic lattice is coherent, similar to Bragg reflection of x-rays, which leads to the Bragg condition that the momentum transferred to the crystal must be a reciprocal lattice vector, $\vec{G} = \frac{2\pi}{a_0}(h,k,l)$ where $a_0$ is the size of the conventional cubic cell and h, k, and l are integers [9]. If we assume the charge distribution in the crystal is simply a superposition of atomic charge distributions, the rate of conversion per unit flux of axions of energy $E_a$ and momentum $\vec{k}_a$ to photons of energy $E_\gamma$ and momentum $\vec{k}_\gamma$ is:

$$\frac{d\dot{N}}{d\Phi}(E_a, \vec{k}_a; E_\gamma, \vec{k}_\gamma) = \frac{V}{v_c^2}(\hbar c)^3 \sum_G |S(G)|^2 \frac{1}{E_a^2}\frac{d\sigma}{d\Omega}(\vec{G})\,\delta(E_a - E_\gamma)\,\delta(\vec{k}_a - \vec{k}_\gamma - \vec{G}) \qquad (6)$$

where V is the volume of the crystal, $v_c$ is the volume of the unit cell, and S(G) is the structure function for the crystal, which for germanium is:

$$S(G) = \left[1 + e^{\frac{i\pi}{2}(h+k+l)}\right]\left[1 + e^{i\pi(h+k)} + e^{i\pi(h+l)} + e^{i\pi(k+l)}\right]. \qquad (7)$$

Table 1 gives the parameters of the most relevant reciprocal lattice vectors of germanium.

If we integrate over all final photon states we find the total conversion rate of axions with momenta $\vec{k}_a$ and energy $E_a$ to be

$$\frac{d\dot{N}}{d\Phi} = 2\hbar c \frac{V}{v_c^2} \sum_G |S(G)|^2 \frac{d\sigma}{d\Omega}(G) \frac{1}{|\vec{G}|^2} \delta\left(E_a - \frac{\hbar c |\vec{G}|^2}{2\vec{k}_a \cdot \vec{G}}\right). \quad (8)$$

Note that in equation (8) the conversion rate is essentially zero unless the energy and momentum of the axion exactly satisfies the Bragg condition. In a real crystal, the outgoing x-ray very probably produces a photo-electron after traveling a distance of order microns which tends to smear out the delta function. However, so long as the mean free path of the x-ray is long compared with the size of the unit cell (as it is in germanium), the broadening of the delta function will always be much less that the actual energy resolution of the detector and can be neglected.

If we now multiply this by the flux of solar axions, (5), the total rate of conversion of axions is,

$$\frac{d\dot{N}}{dE}(\hat{k}, E_a) = 2\hbar c \frac{V}{v_c^2} \sum_G |S(G)|^2 \frac{d\sigma}{d\Omega}(G) \frac{1}{|\vec{G}|^2} \frac{d\Phi}{dE} \delta\left(E_a - \frac{\hbar c |\vec{G}|^2}{2\hat{k} \cdot \vec{G}}\right), \quad (9)$$

where $\hat{k}$ is the unit vector from the sun, and $d\Phi/dE$ is evaluated at the axion energy $E_a = \hbar c |\vec{G}|^2 / 2\hat{k} \cdot \vec{G}$. This can be written in the compact form:

$$\frac{d\dot{N}}{dE}(\hat{k}, \epsilon) = \lambda\, M_d\, \dot{N}_0 \sum_g |S(g)|^2 \left[\frac{4\epsilon^2 - g^2}{g^2 + \gamma^2}\right] \frac{\epsilon^3}{e^{\beta\epsilon} - 1} \delta\left(\epsilon - \epsilon(\vec{g})\right) \quad (10)$$

where $M_d$ is the mass of the detector in kg, $\vec{g} = (h, k, l)$, $\gamma = a_0/r_0$, $\beta = \dfrac{2\pi\hbar c}{a_0 E_0}$, $\epsilon(\vec{g}) = \dfrac{g^2}{2\hat{k} \cdot \vec{g}}$, and $\dot{N}_0 \equiv \dfrac{\sigma_0 \Phi_0 \beta^2}{\rho(2\pi a_0)^3} = 0.61 / \text{kg} \cdot \text{d}$ for germanium.

For a given direction of incident axions, $\hat{k}$, the Bragg condition for each reciprocal lattice vector such that $\hat{k} \cdot \vec{G} > 0$ determines a narrow range of energies. A germanium detector has a characteristic low-energy resolution, $\Delta E \sim 400$ eV. We take this into account by replacing the delta function in (10) by a gaussian with the same fwhm as the detector. Finally we calculate the total counting rate in a range of energies of width $\Delta\epsilon$,

$$R(\hat{k}, \epsilon) = \int_{\epsilon - \Delta\epsilon/2}^{\epsilon + \Delta\epsilon/2} d\epsilon' \frac{dN}{dE}(\hat{k}, \epsilon'). \tag{11}$$

In figure 2 we show the expected counting rate, $R(\hat{k}(t),\epsilon) \equiv R(t)$, as a function of time over a single day. The position of the sun is calculated using the U. S. Naval Observatory Vector Astronomy Subroutines (NOVAS) [10].

The pronounced variation in the axion counting rate as a function of time suggests that events in the detector be analyzed in terms of the correlation function:

$$\chi = \int_0^T [R(t) - \overline{R}] n(t) dt, \tag{12}$$

where R(t) is the theoretical instantaneous axion counting rate in a given energy interval, T is the time during which the detector is on, $\overline{R}$ is the average of R(t) over this period and n(t) is the number of counts at time t in a short time interval $\Delta t$. Note that if n(t) is uncorrelated with the position of the sun, then $\chi \sim 0$ within statistical fluctuations, whereas if n(t) contains an axion component, $\chi$ will increase proportionally to T.

We have carried out Monte Carlo simulations of a germanium detector with realistic background rates taking the range of energies from 2 to 10 keV in 0.5 keV intervals. We find that an array of 400, oriented single crystal n-type detectors, each of 0.25 kg operating for 10 years can set an upper limit on the scale factor $\lambda$,

$$\lambda \leq 10^{-4}. \tag{13}$$

This corresponds to a 95% CL bound on the axion-photon coupling constant $g_{a\gamma\gamma} = 1/Mc^2 \approx 5 \times 10^{-10}$ GeV$^{-1}$ for axion masses between 0 and about 1 keV.

Similar results could be obtained using ~160, single crystals of p-type detectors, each of 0.65 kg. A negative result would exclude a significant portion of the remaining $g_{a\gamma\gamma}$ - $m_a$ phase space. A pilot experiment has been done which has accumulated data for approximately 1.92 kg-years and is being analyzed by the methods discussed above [11]. This result was derived using the correlation function of equation (11). It is not clear that more sensitive analysis techniques employing pattern recognition, for example, might not be more effective. these are under investigation at present.

This large array of low energy threshold, low background detectors with excellent energy resolution would also be a sensitive detector for weakly interacting massive particles hypothesized as cosmic cold dark matter.

The present technique could also be applied to low temperature sapphire detectors planned for dark matter searches, and to tellurium oxide detector arrays being constructed for double-beta and dark matter searches. In these searches, nuclear recoil events are of interest, whereas for an axion search, only events associated with photon signals are relevant. The same is true for large arrays of Ge detectors which also employ cryogenic bolemetric techniques. For these to be more effective than an ordinary low-background Ge detectors, the raw background due to photons would have to be low.

Acknowledgments

This work was supported by the US National Science Foundation Grand INT930INT1522, and the University of South Carolina Summer Institute for the Foundations of Physics. S. Nussinov would like to acknowledge a BSFG (Israel USA Foundation Grant). The authors would like to thank J. A. Bangert of the U. S. Naval Observatory for supplying their vector astronomy subroutines, and Y. Aharonvov and E. A. Paschos for helpful discussions.

Captions for the Figures

Figure 1. Axion conversion to a photon in the Coulomb field of the nucleus by the Primakoff effect.

Figure 2. Theoretical prediction of the count rate of photons converted from axions incident at a Bragg angle, for a detector located at Sierra Grande, Argentina (41º 41′ S, 65º 22′ W). The rate was calculated for $1/Mc^2 = g_{a\gamma\gamma} = 10^{-8}$ GeV$^{-1}$. The location that was chosen is where the pilot experiment is being performed.